\documentclass[aps,prl,twocolumn,superscriptaddress,groupedaddress,a4paper,showpacs]{revtex4-1} 
\usepackage{graphicx}
\usepackage{dcolumn}
\usepackage{bm}

\usepackage{color}

\begin{document}


\title{Anomalous features of diffusion in corrugated potentials with spatial  correlations:
faster than normal, and other surprises.}

\author{Igor Goychuk}
 \email{igoychuk@uni-potsdam.de, corresponding author}
 
\affiliation{Institute for Physics and Astronomy, University of Potsdam, 
Karl-Liebknecht-Str. 24/25,
14476 Potsdam-Golm, Germany}

\author{Vasyl O. Kharchenko}

\affiliation{Institute of Applied Physics, 58 Petropavlovskaya str., 40030 Sumy, Ukraine}

\date{\today}

\begin{abstract}

Normal diffusion in corrugated potentials with spatially 
uncorrelated Gaussian energy
disorder famously explains the origin of non-Arrhenius $\exp[-\sigma^2/(k_BT^2)]$ 
temperature-dependence in
disordered systems. Here we show that unbiased diffusion remains asymptotically normal also in 
the presence of spatial correlations decaying to zero. However, 
due to a temporal lack of self-averaging
transient subdiffusion emerges on mesoscale, and it can readily reach macroscale
even for moderately strong disorder fluctuations of $\sigma\sim 4-5\, k_BT$. 
Due to its nonergodic origin
such subdiffusion exhibits a large scatter in single trajectory averages. However,
at odds with intuition, it occurs essentially faster than one expects
from the normal diffusion in the absence of correlations.
We apply these results to diffusion of regulatory proteins on DNA molecules
and predict that such diffusion should be anomalous, but much faster than earlier expected 
on a typical length of  genes for a realistic energy disorder of several room $k_BT$,
or merely $0.05-0.075$ eV. 

\end{abstract}

\maketitle

Diffusion and transport processes in disordered amorphous materials, including various 
polymer glasses and biopolymers such as DNAs and proteins are in the research limelight 
already for over fifty years \cite{Bouchaud,HTB90,Hughes}. A paradigm in this field is provided 
by hopping transport 
modeled  by continuous time random walks (CTRW)  with energy disorder on the sites of 
localization and their  continuous space analogy, where the continuous space diffusion of 
overdamped particles
is considered in random potentials (static or quenched disorder). 
Exponential energy disorder on sites 
can easily yield anomalous diffusion and transport when the dispersion of energy fluctuations 
$\sigma$ exceeds thermal energy $k_BT$. Such an exponential model of energy disorder 
uncorrelated on sites  gave rise
to  famous CTRW model of anomalous transport by Montroll, Scher, Weiss, Shlessinger, and others
 \cite{Hughes,Metzler}. It is
featured by heavy-tailed residence time distributions (RTDs) on sites, 
$\psi(\tau)\sim \tau^{-1-\alpha}$, possessing no mean value, 
with $\alpha \sim k_BT/\sigma$ within
a mean-field approximation. However, strictly exponential energy disorder does not yield any of the well-known 
non-Arrhenius 
temperature dependences of diffusion coefficient $D$ and mobility $\mu$ in glass-like materials such as
$D(T)\propto \exp[-\sigma^2/(k_BT)^2]$ \cite{HTB90,DeGennes,Zwanzig,Bassler87,Bassler93}, 
or the Vogel-Fulcher 
law
$D(T)\propto \exp[-\sigma/(k_B(T-T_0)]$, for $T>T_0$ \cite{Vilgis}, which are not easy to
distinguish experimentally \cite{Hecksher}. 
On the other hand, the model of Gaussian, rather
than exponential
energy disorder has been justified for a number of materials \cite{Bassler93}. 
Gaussian disorder emerges naturally  by virtue of the central limit theorem e.g. in 
molecularly doped polymers with dipolar disorder \cite{Dunlap}.
Furthermore,
genetic material has been foreseen as an aperiodic disordered crystal already by 
Schr\"odinger in his famous ``What is life?''\cite{Schrodinger}.
Indeed, interaction of transcription factors and 
signaling proteins with DNA macromolecules -- a problem central for gene expression
in molecular biology -- is also well described by the Gaussian energy disorder \cite{Lassig,SlutskyPRE}. 
If Gaussian disorder
is spatially uncorrelated, no anomalous diffusion and transport regime is possible.
This is because any Gaussian energy disorder yields
in the mean-field approximation local RTDs with all the moments being finite. 
Accordingly, the classical
result by de Gennes, Zwanzig, and B\"assler yields the renormalization (suppression) of normal transport
coefficients by the factor $\exp[-\sigma^2/(k_BT)^2]$. This famously explains the origin of this
remarkable temperature dependence \cite{DeGennes,Zwanzig,Bassler87,HTB90}.
However, in dipolar organic glasses the long range 
correlations in site energy fluctuations emerge \cite{Dunlap} . 
Short range correlations also naturally emerge for diffusion of proteins 
on DNAs.
Indeed, let us consider a contact area of DNA and a bound protein. It involves typically from 5 to 30 
base pairs (bp) in length \cite{Stewart}.
The interaction energy is a pairwise sum of the energy of interaction of a base in contact and protein. 
It is approximately Gaussian distributed \cite{Lassig}. 
When protein slides by one base 
along DNA, 
it remains in contact with all the same bases except one new and one past. This
fact most obviously introduces spatial correlations in the random binding energy profile on a typical
length of DNA-protein contact, even if pairwise correlations are 
totally absent. 
Obviously, any correlations in the bp sequence or inclusion of long-range electrostatic interactions 
\cite{Cherstvy} can only enhance spatial range of such correlations.
This provokes the question: How do the binding energy correlations affect
diffusion along DNA?
Will it be still normal, or
maybe anomalous diffusion regime emerges? Below we show,  
that if spatial correlations decay
to zero, diffusion is asymptotically normal.
Vanishing of spatial correlations
guarantees self-averaged ergodic character of unbiased diffusion
on very large distances.
Renormalized 
diffusion coefficient is  described by the same well-known 
result of Ref. \cite{Zwanzig}. 
However, some older \cite{Romero}  and very recent \cite{Khoury,Simon} simulations 
do reveal anomalous diffusion and transport. 
Is something wrong with these simulations? 
No, we confirm them in some basic features. Anomalous diffusion  emerges indeed.
However, contrary to the earlier argumentation  \cite{Khoury} it is not
based on a residence time distribution with divergent moments.
Subdiffusion can last very long because on the corresponding mesoscale no self-averaging is attainable.
However, on very large distances  
it smoothly changes into normal diffusion.
This provokes the question: "How large is very large?" What determines 
the corresponding mesoscale?
When the classical result is indeed
physically relevant, and when it becomes of lesser utility, or can even mislead?
These are the major questions we answer with this work. 

As a most striking result,
spatial correlations not only introduce subdiffusion, but this subdiffusion
proceeds much faster than 
expected from 
exponentially suppressed
normal diffusion.
Averaged exit times from any finite spatial domain
and their variance are not only finite,
but they become much  smaller than 
in the absence of correlations.
Spatial disorder correlations lead to transient subdiffusion. 
However, this transient subdiffusion makes mesoscopic transport processes overall 
faster, not slower, as
generally believed.
This important result conforms
to previous conclusions in \cite{Goychuk12} obtained within other
modeling frameworks.

\begin{figure}
\includegraphics[width=8cm]{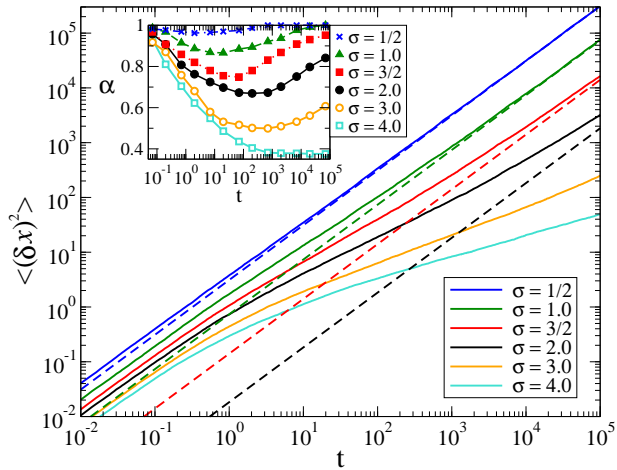}
\caption{(Color online) Ensemble-averaged diffusion for different values of
disorder strength $\sigma$ in units of  $k_BT$ for exponentially decaying correlations. In doing numerics, we
fixed $\sigma=\sigma_0$ and varied temperature. Distance is measured in units 
of correlation length $\lambda$ and time in units of $\tau_0=\lambda^2\eta/\sigma_0$. 
For $\sigma_0=2\, k_BT_{\rm room}=0.05$ eV,
$D_0=k_BT_{\rm room}/\eta=3\;\mu m^2/s$, and $\lambda=5.25$ nm (15 bp), $\tau_0\approx 4.6\;\mu s$.
 Initially, diffusion is normal, 
$\langle \delta x^2(t)\rangle =2 D_0 t$.
The dashed lines present asymptotically normal behavior 
$\langle \delta x^2(t)\rangle =2 D_{\rm ren}t$, for $\sigma/(k_BT)=1/2,1,3/2,2$. 
Transient subdiffusion is much faster than this limit!
Averaging
over $10^4$ particles is done in 10 different realizations of random potential replicated with
period $L=10^4$ ($10^3$ particles per a potential realization). Particles are initially uniformly 
distributed over the length $L$.
}
\label{Fig1}
\end{figure}

\textit{Model.}
We consider a standard model of overdamped 
diffusion 
in a spatially disordered potential
$V(x)$ \cite{HTB90,DeGennes,Zwanzig}. It is described by 
Langevin equation
\begin{eqnarray}
 \eta \dot x=-\frac{\partial U(x)}{\partial x}+\sqrt{2k_BT\eta}\zeta(t),
\end{eqnarray}
at temperature $T$. Here, $\eta$ is frictional coefficient, and $\zeta(t)$ is unbiased white Gaussian 
noise, $\langle \zeta(t)\zeta(t')\rangle=\delta(t-t')$. The potential energy, $U(x)=U_{\rm reg}(x)+V(x)$, 
consists generally
 of two parts, regular $U_{\rm reg}(x)$, e. g. $U_{\rm reg}(x)=-f_0 x$ 
 for a constant force $f_0$, and a random part $V(x)$.  
 It obeys  unbiased Gaussian 
distribution, $\langle V(x)\rangle=0$, 
with variance $\sigma^2$, and
normalized correlation function $g(z)$,
\begin{eqnarray}
\langle V(x)V(x')\rangle =\sigma^2 g(|x-x'|) \;,
\end{eqnarray}
$g(0)=1\geq g(z)$, being a wide sense stationary random process in space. In application
to diffusion on DNA, regular potential includes also 
a mean binding energy $V_0\sim 10-20\;k_BT_{\rm
room}$, and
$\sigma\ll |V_0|$. $V_0$ is crucial for the protein binding and dissociation, but it 
does not influence sliding along DNA. A simplest model is provided by
exponentially decaying short-range correlations, $g(z)=\exp(-|z|/\lambda)$, with correlation 
length $\lambda$, which is about the linear size of protein-DNA contact. 
In numerical simulations, 
this model 
was effectively regularized to make the mean square fluctuation of random force 
$f(x)=-\frac{\partial V(x)}{\partial x}$ finite 
\cite{Supplement},

\textit{Theory and Results.} 
Normal transport coefficients renormalized by disorder can be found by a standard
trick with periodization of random potential \cite{Bouchaud}, 
imposing an artificial spatial
period $L$, and considering the limit $L\to\infty$ at the end of calculation. Following Refs.
\cite{Zwanzig,Dunlap}, one obtains (at finite $L$)
\begin{eqnarray}\label{diffusion}
D_{\rm ren}= 
\frac{D_0}{\overline{C_L^+}\; \overline{C_L^-} }\;
\end{eqnarray}
for the renormalized diffusion coefficient in the unbiased case ($f_0\to 0$)\cite{Supplement}.
Here,
\begin{eqnarray}\label{corr-fun}
\overline{C_L^{\pm}} = \frac{1}{L}\int_0^{L} e ^{\pm\beta V(x)}dx
\end{eqnarray}
is a spatially averaged random function $w_{\pm}(x):=e ^{\pm\beta V(x)}$. Furthermore, 
$D_0=k_BT/\eta$ is free diffusion coefficient, and 
$\beta=1/(k_BT)$ is inverse temperature.
The earlier result \cite{Zwanzig,Parris} readily follows upon identifying the spatial
average in Eq. (\ref{corr-fun}) with the ensemble average
$\langle C_L^{\pm}\rangle  = \langle e ^{\pm \beta V(x)}\rangle$ over random
realizations of $V(x)$.
Since for any zero-mean Gaussian variable $\xi$,
$\langle \exp(\xi)\rangle=\exp (\langle \xi^2\rangle/2)$, 
for arbitrary Gaussian disorder 
we obtain 
 \begin{eqnarray}\label{diffusion2}
D_{\rm ren}= D_0 e^{-\sigma^2/(k_BT)^2}\;.
\end{eqnarray}
Remarkably, this is precisely the same result obtained earlier for non-correlated 
potentials \cite{Zwanzig}. 
Correlations make no influence on it! 
This is a very important conclusion and numerics
completely confirm it in Fig. \ref{Fig1}, for the particular model considered.

Here comes our crucial point. Namely, we wish to reexamine the ergodic assumption
leading to (\ref{diffusion2}). When does it work
in the strict limit $L\to\infty$? Even more important, for which \textit{finite} $L$
it becomes well justified? This will give us 
a characteristic mesoscopic scale of transiently anomalous  diffusion.
For $L$ smaller than a typical ergodicity length $L_{\rm erg}$ we expect anomalous 
diffusion, which becomes asymptotically normal for $L\gg L_{\rm erg}$.
To establish the corresponding criterion one has to consider statistical variations
of $\overline{C_L^{\pm}}$. 
Following a standard procedure \cite{Papoulis}, we consider the (relative) ensemble variance,
$\left [\left \langle  \left (\overline{C_L^{\pm}}\right)^2\right \rangle -
\left \langle \overline{C_L^{\pm}}\right \rangle^2\right]/
\left\langle \overline{C_L^{\pm}}\right\rangle^2$,
of the 
trajectory average $\overline{C_L^{\pm}}$ named also ergodicity breaking parameter (EBP) 
\cite{Deng}. It must vanish
for any ergodic process in the limit $L\to\infty$. Then, one can  use 
$\langle C_L^{\pm}\rangle$ instead of $\overline{C_L^{\pm}}$. 
Sufficient
condition for this is  that  the ensemble-averaged autocorrelation function
$K_{\pm}(x)=\langle \delta w_{\pm}(x_0)\delta w_{\pm}(x_0+x)\rangle$ of the random
process $\delta w_{\pm}(x):=e ^{\pm\beta V(x)}-\langle e ^{\pm\beta V(x)} \rangle$
vanishes in the limit $x\to \infty$ \cite{Papoulis}.  After some straightforward algebra we obtain
\begin{eqnarray}\label{Kxy}
K_{\pm}(x)=e^{\beta^2\sigma^2}\{\exp[\beta^2\sigma^2 g(x)] -1\}\;.
\end{eqnarray}
From this important result it follows immediately, that diffusion
is indeed asymptotically ergodic and normal for any random Gaussian potential with vanishing
correlations,
$\lim_{x\to \infty }g(x)=0$. Then 
the result in 
Eq. (\ref{diffusion2}) is valid.

We focus on short-ranged correlations,
which seemingly justified the
use of the approximation of uncorrelated disorder in the bulk of previous
research work \cite{DeGennes,Zwanzig,HTB90}. Even here, with growing $\sigma$ diffusion 
becomes transiently anomalous,
$\langle \delta x^2(t)\rangle \propto t^{\alpha(t)}$, with a time-dependent 
$0<\alpha(t)\leq 1$. It starts from $\alpha=1$ at $t=0$ and tends
to $\alpha=1$ asymptotically, see inset in Fig. \ref{Fig1}. The time duration and spatial 
extension of subdiffusion depend very strongly on $\sigma$. For example, for $\sigma=4$
in Fig. \ref{Fig1}, there is no any signature of growing $\alpha(t)$ on the whole
time scale of simulation. Indeed, $\alpha\approx 0.4$ for $10^3<t<10^5$.
The emergence of this subdiffusion is due to a transient breaking of ergodicity.
Importantly, it is also 
non-Gaussian in subdiffusive regime,
see Fig. S2 in \cite{Supplement}.
There exists an ergodicity length $L_{\rm erg}(\sigma)$, such that self-averaging occurs
only for $L\gg L_{\rm erg}(\sigma)$. However, no self-averaging occurs on the
mesoscale 
defined by the requirement that the above EBP 
equals one, which leads to the condition
\begin{eqnarray} \label{cond}
\int_0^1\left (1-y\right )e^{\beta^2\sigma^2g(L y)}dy=1 \;.
\end{eqnarray}
Solution of this equation for unknown $L$ gives $L_{\rm erg}$.
Another estimation yields $L_{\rm erg}(\sigma)
\sim \lambda e^{\sigma^2/(k_BT)^2}$ \cite{SlutskyPRE}, which indeed displays a major 
trend with $\sigma$. For example, for $\sigma=2$, 
Eq. (\ref{cond}) yields $L_{\rm erg}(2)\approx 35\lambda$ 
(while $e^4\approx 54.6$).  This indeed is consistent with the 
trend one observes in Fig. 
\ref{Fig1} for $\sigma=2$,  where $\langle \delta x^2(t_{\rm max})\rangle \sim 3000 \lambda^2$.
In this respect,
a recent experiment shows that 1d diffusion
along DNA is suppressed by the factor of hundred with respect to one in the
bulk \cite{Elf}. This suggests  $\sigma\sim 2\, k_BT_{\rm room}\sim 0.05$ eV with
experimental values $D_0=3\;\mu m^2/s$ and $D_{\rm ren}=0.046\;\mu m^2/s$ \cite{Elf}.
Applying this result to diffusion of
a protein on DNA with $\lambda=15$ bp suggests that protein diffusion should
be still anomalous on a typical gene length about 1000 bp. Remarkably,  another
experiment reveals even a larger suppression factor of about $10^4$ \cite{Wang}, which
would correspond to $\sigma\sim 3\;k_BT_{\rm room}$. Then, from Eq. (\ref{cond})
 $L_{\rm erg}(3)\approx 2070\;\lambda$ (while $e^9\approx 8103 $), a drastic increase! 
 This would mean that a typical subdiffusion length would cover about 30-120 genes in
 bacterial genome, which can have important consequences for gene regulation.
  In a more general context, already for $\sigma=4\, k_BT$,
 $L_{\rm erg}(4)\sim 1.2\cdot 10^{6} \lambda$, i.e. for  $\lambda\sim 10$\AA ,
 $L_{\rm erg}(4)\sim 1.2$ mm, i.e. subdiffusion reaches clearly a macroscale.
 Other models of decaying correlations cannot change much this conclusion. Then, the
 classical result in Eq. (\ref{diffusion2}) can mislead, even being formally valid.

\begin{figure}
\includegraphics[width=8cm]{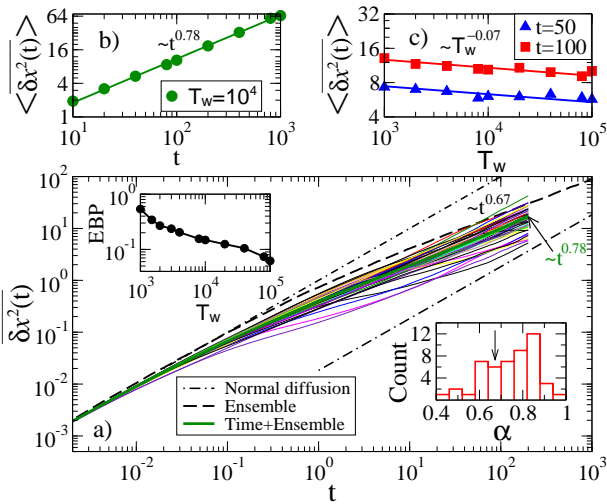}
\caption{(Color online) (a) Single-trajectory averages are scattered between the free
and disorder-renormalized diffusion
limits (depicted with dashed-dotted lines).
Time window ${\rm T_w}=2\cdot 10^4$ for averaging
is chosen 100x larger than the maximal time $t$. Each trajectory is characterized
by individual subdiffusion index $\alpha$ distributed as shown in the lower inset. Arrow indicates
the value of $\alpha=0.67$ which corresponds to the ensemble average depicted with dashed line. The ensemble 
average of time-averages  with $\alpha=0.78$ is depicted as full green line. It is also depicted 
in part (b) as function of time $t$ for ${\rm T_w}=10^4$, and also in part (c) as function of 
${\rm T_w}$ for two fixed values of $t$. The latter one decays as  $\Big \langle
 \overline{ \delta x^2(t)}\Big\rangle \sim {\rm T_w}^{-0.07}$. Moreover, the corresponding 
 ergodicity
 breaking parameter in the upper inset of part (a) gradually decays as a function of 
 ${\rm T_w}$.  This indicates that no ergodicity breaking takes place asymptotically, 
 ${\rm T_w}\to\infty$. 
}
\label{Fig2}
\end{figure} 

\begin{figure}
\includegraphics[width=8cm]{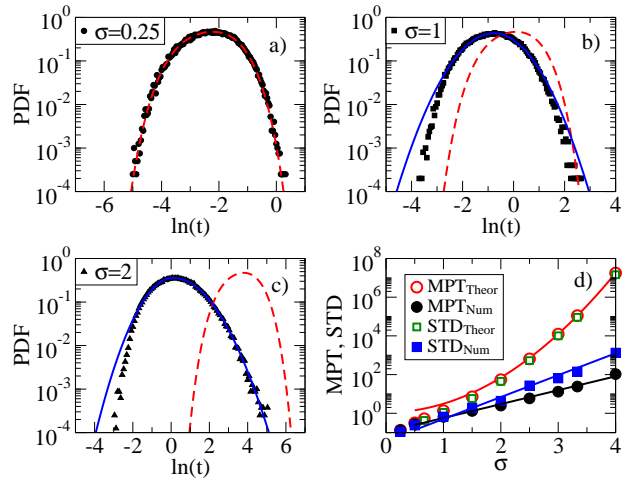}
\caption{(Color online)(a,b,c) Probability density of waiting times derived from 
numerics (symbols), its fit with a 
log-normal distribution (full lines), and the result expected from normal diffusion
with $D_{\rm ren}$ (dashed lines) for three different values of $\sigma/k_BT$, (d) Mean first
passage time (MPT) and its standard deviation (STD) derived from numerics (symbols) and their
exponential fits (full lines), as well as MPT and STD expected from normal diffusion characterized
by $D_{\rm ren}$, which obey  a $const\times\exp[-\sigma^2/(k_BT)^2]$ 
dependence.
}
\label{Fig3}
\end{figure}  
 
Given nonergodic origin of such subdiffusion it becomes  important 
to study single-trajectory averages, 
$\overline{ \delta x^2(t)}^{\rm T_w}:=[1/({\rm T_w}-t)]\int_0^{{\rm T_w}-t}[\delta x(t|t')]^2dt'$,
of the mean-squared
displacement, $\delta x(t|t')=x(t+t')-x(t')$, over a time window, ${\rm T_w}$,  
 assuming $t\ll {\rm T_w}$ \cite{Wang,Golding}. The results for $\sigma=2$ 
 display a typical scatter in Fig. \ref{Fig2}. It can be characterized by a broadly distributed 
 subdiffusion exponent $\alpha$. Similar features are indeed seen in many
 experiments \cite{Golding}. 
 The corresponding ensemble average $\Big \langle
 \overline{ \delta x^2(t)}^{\rm T_w}\Big\rangle$ is different from the standard ensemble average 
 $\langle \delta x^2(t)\rangle$, having even a different anomalous exponent, see Fig. \ref{Fig2}, a, b.
 Recent experimental findings \cite{Wang} indirectly corroborate our results. Indeed,
 in  \cite{Wang} a huge scatter of the diffusional constants for LacI protein on a bacterial
 DNA has been reported, which the authors attributed to a wildly (over 3 orders of magnitude)
 distributed normal diffusion coefficient. When we increase $\sigma$ to 
 $\sigma=3\;k_BT$, the scatter indeed further increases, see in Fig. S3 of \cite{Supplement}.
 
 Strikingly enough, all the single-trajectory averages reveal  subdiffusion which proceeds 
 much faster than expected from Eq. (\ref{diffusion2}),
 see in Fig. \ref{Fig2}, a. 
 It must be emphasized that even though our results somewhat remind ones obtained for CTRW subdiffusion 
 with divergent mean 
 residence times, or with exponential energy disorder \cite{CTRWage}, in fact, they are very different.
 First, also single trajectory averages yield subdiffusion (without any boundary effects). Second,
 the drift of these averages with growing time window ${\rm T_w}$ is much less pronounced. 
 see in Fig. \ref{Fig2}, c. Moreover,  the related EBP shows 
 a clear tendency to zero with increasing ${\rm T_w}$, cf. inset
 in Fig. \ref{Fig2}, a. 
 
 Especially important is that the corresponding residence time distribution
 to stay in any finite-size spatial domain 
 is neither featured by diverging mean residence time, nor by diverging variance. 
 In this respect, our results also essentially differ
 from the results in \cite{Khoury}. They are somewhat closer in this particular 
 aspect to viscoelastic subdiffusion.
 However,  the
 latter one is mostly ergodic by its origin \cite{Goychuk09}, and therefore is also different.
  We investigate the distribution of escape
 times out of spatial domain 
 $[-\lambda,\lambda]$ for the particles initially localized 
 in the middle of it. For  disorder-renormalized  normal 
 diffusion the residence time distribution  can be 
 derived as 
 $\psi(t)=\pi\sum_{n=0}^{\infty}(-1)^n(2n+1)e^{-\pi^2(2n+1)^2t/4}$, with time in units
 of $\lambda^2/D_{\rm ren}$. It
 is dominated by a single-exponential $\psi(t)\propto \pi \exp(-\pi^2 t/4)$ at large times.
  For small disorder,
 this result is nicely confirmed numerically in
 Fig. \ref{Fig3}, a, which provides
 also one of the successful tests of the accuracy of our numerics. 
 However, already for $\sigma=k_BT$ essentially deviations are
 observed in Fig. \ref{Fig3}, b. The mean time not only exists, but it is much smaller
 that one expects from normal diffusion, even though the distribution becomes broader 
 than exponential. For a sufficiently large disorder, 
 its essential part is nicely described by the log-normal distribution, 
 $\psi(t)=1/(\sqrt{2\pi}\sigma_\tau t)\exp[-\ln^2(t/\tau_0)/(2\sigma_\tau^2)]$,
 with two parameters $\tau_0$ and $\sigma_\tau$,  
 which are related to the finite mean and variance of this distribution depicted in Fig. \ref{Fig3}, d.  
 Such a distribution can be confused  for a power law distribution, 
  $\psi(t)\propto 1/t$, at short times. However, it is profoundly different.
  The numerical mean and variance are much smaller than those
expected from disorder-renormalized normal diffusion. Moreover, they exhibit
a linear dependence on $\sigma/(k_BT)$ in the exponential, i.e. $\propto \exp[\sigma/(k_BT)]$, rather
than quadratic, i.e.   $\propto \exp[\sigma^2/(k_BT)^2]$. Fig. \ref{Fig3}, d
illustrates this very important finding. Subdiffusional search due to spatial correlations
is thus expected to proceed much faster that one naively expects from the well-known 
renormalization by disorder! 

To conclude, we summarize the important findings of this work. First, the 
famous result in Eq. (\ref{diffusion2}) remains valid asymptotically for any model of
decaying correlations. Diffusion  is suppressed 
by the factor responsible for well-known non-Arrhenius temperature dependence \cite{Hecksher}.
 However, a similar factor  characterizes also the spatial
range of transient subdiffusion in units of the disorder correlation length $\lambda$.
Second, subdiffusion readily reaches a macroscale even for
a moderately strong disorder of $\sigma \sim 4-5\, k_BT$.
Third, already for
$\sigma\sim 2\, k_BT_{\rm room}\sim 0.05$ eV diffusion of regulatory proteins on DNAs becomes
essentially anomalous on a typical length of genes, with a large scatter in single trajectory averages.
Fourth, and the most surprising, such subdiffusion proceeds much faster
than one expects when applies
Eq. (\ref{diffusion2})
to transport processes on mesoscale.
We believe that these important findings provide a new vista on the role of correlations in
Gaussian disorder and  subdiffusion, and will inspire related experimental
work.

\textit{Acknowledgment.} 
Support of this research by the Deutsche Forschungsgemeinschaft (German Research Foundation), Grant
GO 2052/1-2 is gratefully acknowledged.

\end{document}